\documentclass[%
 aip,
rsi,%
 amsmath,amssymb,
 reprint,%
]{revtex4-1}

\usepackage{graphicx}
\usepackage{dcolumn}
\usepackage{bm}

\begin{document}

\title{Developing of a photonic hardware platform for brain-inspired computing  based on $5\times5$ VCSEL arrays}

\author{T. Heuser}
\affiliation{Institut für Festk\"{o}rperphysik, Technische Universit\"{a}t Berlin, Hardenbergstraße 36, 10623 Berlin, Germany.}%

\author{M. Pfl\"{u}ger}
\affiliation{Instituto de F\'{i}sica Interdisciplinar y Sistemas Complejos, IFISC (UIB-CSIC), Campus Universitat de les Illes Baleares, E-07122 Palma de Mallorca, Spain}%

\author{I. Fischer}
\affiliation{Instituto de F\'{i}sica Interdisciplinar y Sistemas Complejos, IFISC (UIB-CSIC), Campus Universitat de les Illes Baleares, E-07122 Palma de Mallorca, Spain}%

\author{J. A. Lott}
\affiliation{Institut für Festk\"{o}rperphysik, Technische Universit\"{a}t Berlin, Hardenbergstraße 36, 10623 Berlin, Germany.}%

\author{D. B}
\affiliation{D\'{e}partement d'Optique P. M. Duffieux, Institut FEMTO-ST,  Universit\'e Bourgogne-Franche-Comt\'e CNRS UMR 6174, Besan\c{c}on, France.}%

\author{S. Reitzenstein}
\email{stephan.reitzenstein@physik.tu-berlin.de}
\affiliation{Institut für Festk\"{o}rperphysik, Technische Universit\"{a}t Berlin, Hardenbergstraße 36, 10623 Berlin, Germany.}%

\date{\today}

\begin{abstract}

Brain-inspired computing concepts like artificial neural networks have become promising alternatives to classical von Neumann computer architectures. Photonic neural networks target the realizations of neurons, network connections and potentially learning in photonic substrates. Here, we report the development of a nanophotonic hardware platform of fast and energy-efficient photonic neurons via arrays of high-quality vertical cavity surface emitting lasers (VCSELs). The developed  $5\times5$ VCSEL arrays provide high optical injection locking efficiency through homogeneous fabrication combined with individual control over the laser wavelengths. Injection locking is crucial for the reliable processing of information in VCSEL-based photonic neurons, and we demonstrate the suitability of the VCSEL arrays by injection locking measurements and current-induced spectral fine-tuning. We find that our investigated array can readily be tuned to the required spectral homogeneity, and as such show that VCSEL arrays based on our technology can act as highly energy efficient and ultra-fast photonic neurons for next generation photonic neural networks.
Combined with fully parallel photonic networks our substrates are promising for ultra-fast operation reaching 10s of GHz bandwidths, and we show that a nonlinear transformation based on our lasers will consume only about 100 fJ per VCSEL, which is highly competitive, compared to other platforms.
\end{abstract}

\maketitle

\section{Introduction:}

Artificial neural networks and brain-inspired machine learning concepts have become highly attractive alternatives to classical computing based on conventional von Neumann architectures. These concepts aim at implementing functionalities for complex computational tasks such as fast pattern recognition\,\cite{Brunner.2013}\cite{LeCun.2015}\cite{Silver.2016}and real-time learning\,\cite{Antonik.2017}. Applications include for instance autonomous driving~\cite{Wang.2020}\cite{Wei.2020}, big data analytics~\cite{Li.2020} and the prediction of chaotic systems~\cite{Li.2012}\cite{Yu.2020}. At the same time the demand for information processing with these concepts rises, and software-based neuromorphic solutions implemented on classical computation hardware might soon reach their limits in terms of energy efficiency, speed and latency. Therefore, the development of energy efficient hardware platforms optimized for brain-inspired computing is of crucial importance to support further progress in the field of machine learning and to enable advanced applications that are out of reach using present day computer architectures.

A neural network comprises two fundamental ingredients, neurons and network connections. Neurons, used in a wider sense, nonlinearly map their input information onto their output, and as such numerous nonlinear photonic components such as semiconductor lasers \cite{Brunner.2013}\cite{Bueno.2017}\cite{Hamerly2019}\cite{Heuser.2020}, Mach-Zehnder modulators \cite{Larger2012}\cite{Paquot2012}, saturable absorbers \cite{Dejonckheere2014} or plasmonic devices \cite{Miscuglio2018} have been proposed and experimentally studied. The promises associated with photonic neural networks are that these components readily enable data-processing rates exceeding 10 GSamples/s \cite{Fischer.2016}, but even more that they can directly be interfaced with fully parallel photonic networks. Such parallel photonic networks typically leverage diffraction \cite{Brunner.2015}\cite{Bueno.2018}\cite{Lin2018}, reconfigurable unitary matrices with Mach-Zehnder arrays \cite{Shen2016}\cite{Pai2019} and recently, novel 3D-printed waveguides \cite{Moughames.2019} or continuous diffraction inside a tailored volume \cite{Khoram2019}. Combined, such high performance photonic neurons and networks therefore promise orders of magnitude speed and latency improvement and potentially an abolishment of the significant energy-overheads caused by serial neural network circuitry. High-performance, reliable photonic neurons building upon mature technology such as the surface emitting semiconductor lasers we report here are therefore key in the development of next generation photonic neural networks.

\begin{figure}[!b]
	\begin{center}
		\includegraphics[scale=0.13]{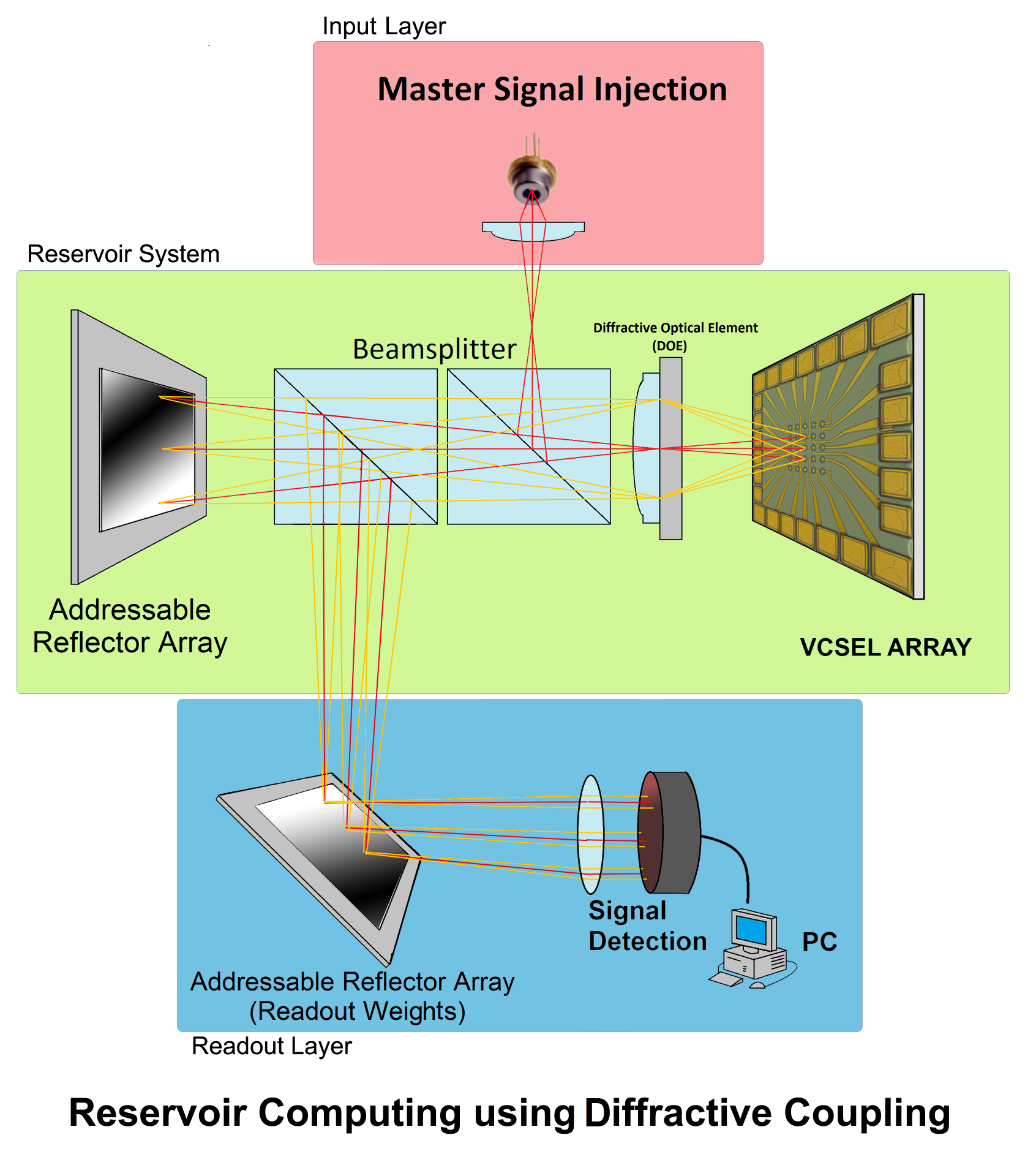}
	\end{center}
	\caption{(a) Schematic presentation of photonic reservoir computing implemented by the diffractive coupling of VCSELs in a dense array, adopted from \cite{Brunner.2015}.}\label{fig1}
\end{figure}

In recent years the reservoir computing (RC) concept \cite{Jaeger.2004} was one of the neural network concepts accelerating hardware implementations with demonstrations in mechanical \cite{Dion.2018}, electrical~\cite{Yi.2016} and photonic~\cite{vanderSande.2017} substrates, where particular photonic implementations promise high computation speed and low energy consumption. One photonic implementation used commercial vertical cavity surface emitting lasers (VCSELs) \cite{Brunner.2015}, but it was found that their relatively low spectral homogeneity posed fundamental limitations. An alternative approach envisioned arrays of quantum dot micropillar lasers~\cite{Heuser.2020}, whose ultra-dense arrays of hosting hundreds of laser neurons within less than 0.5~mm$^2$ is particularly appealing. However, they usually operate under optical pumping at cryogenic temperatures, which hinders real-world applications. In contrast, VCSEL arrays are of high practical relevance, but upscaling the arrays to more than a few tens of lasers will be challenging. The different approaches of laser-based photonic neural networks have in common that the modulated input signal of an injection laser is fed into an array of lasers which are mutually connected via diffractive coupling \cite{Brunner.2015}\cite{Maktoobi.2020}. In this concept a diffractive optical element (DOE) spatially multiplexes each laser precisely at the positions of its neighbors within the array with the help of an external cavity, as shown schematically in Fig.~\ref{fig1} (a). The output of the system can then be calculated from the weighted sum of the lasers' intensities. This kind of RC system is trainable by adjusting the readout weights forming the optical output signal\,\cite{Bueno.2018}\cite{Andreoli.2020}. Once such a  system is implemented, it promises to perform machine learning tasks at GHz speed, where the ultimatelly bandwith limiting factors are the characteristic dynamical time scales of the lasers, typically on the order of 10 ps. From a technological point of view, the dependence of laser-based photonic neural networks on injection locking sets very strict hardware requirements to the laser array in terms of high spectral homogeneity, polarization alignment and geometric dimensions.

Photonic RC based on the diffractive coupling of VCSELs was realized for the first time in~\cite{Brunner.2015}\cite{Bueno.2018} and has proven to be indeed a promising and scalable concept of machine learning. However, the first implementation was based on commercially available, non-optimized VCSEL arrays which limited the performance of the neural network. Here we report on the development of customized  $5\times5$ VCSEL arrays specifically designed and optimized for the needs of photonic RC. Our realization of a  $5\times5$ VCSEL array enables individual biasing of each laser while still keeping a small pitch of around 80\,$\mu$m, which is a factor of 3 smaller than in the individual addressable commercial VCSEL arrays \cite{Brunner.2015}. Furthermore, we realized very high and individually tunable spectral homogeneity as well as a well-aligned and stable polarisation characteristic due to a slightly elliptical cross section of the laser's aperture. 

The article is organized as follows. We first describe technological details of the sample fabrication followed by presenting the overall laser performance. Then we discuss details of the performance and limitations of the injection locking characteristics. We close with a demonstration of the spectral tuning capabilities of the individual lasers and an outlook to future work.

\section{Sample Design and Technology}
\begin{figure*}[!t]
	\begin{center}
		\includegraphics[scale=0.2]{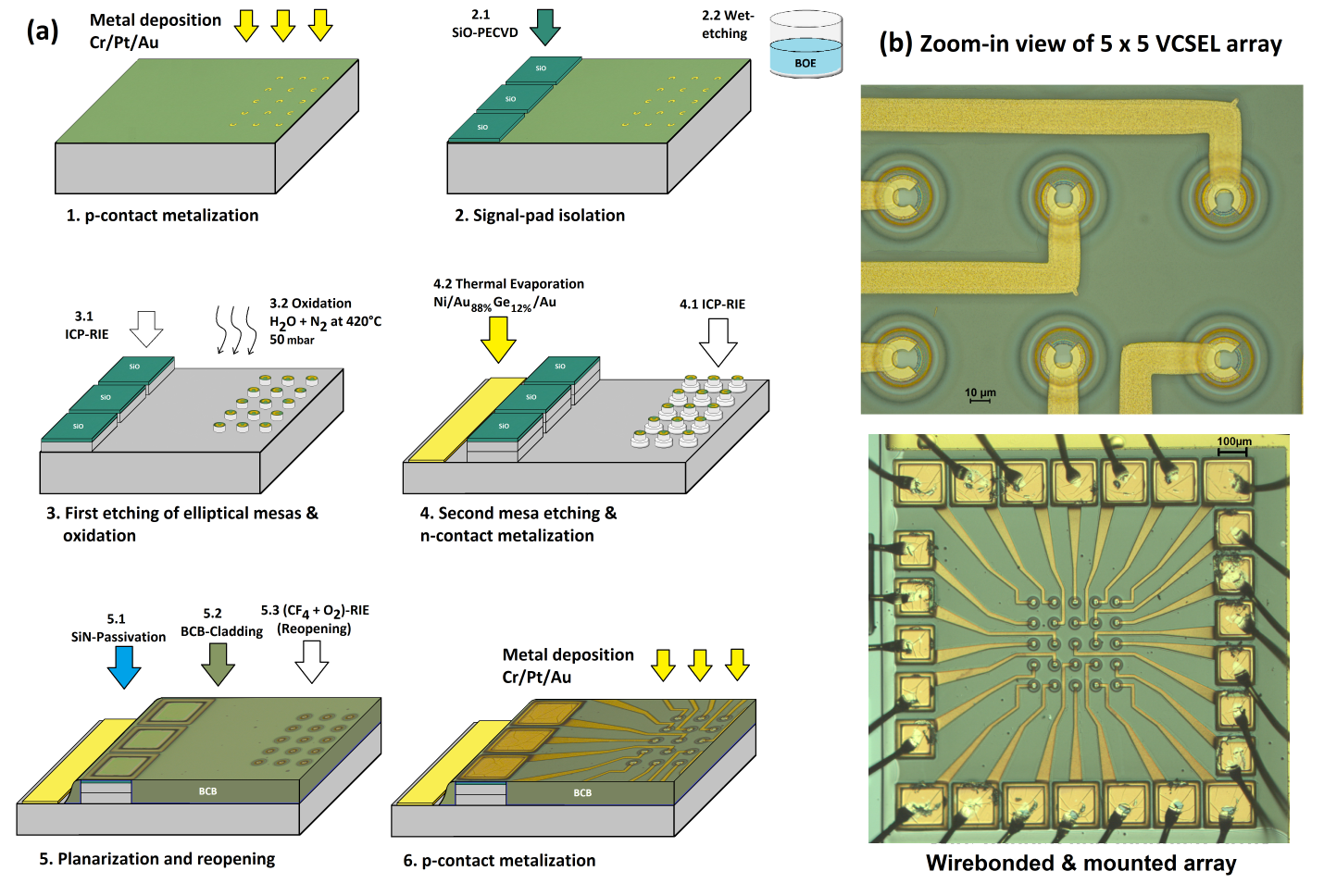}
	\end{center}
	\caption{(a) Schematic overview of the sample fabrication process as described in section 2. (b) Microscope images of the finished VCSELs and the fully wirebonded and mounted array.}\label{fig2}
\end{figure*}
The $5\times5$ VCSEL arrays are designed with a small pitch of around 80\,$\mu$m to meet the maximal dimension of the RC optics field of view with an area of approximately 1 $\mathrm{mm}^2$\,\cite{Maktoobi.2020}. For the given emission wavelength of the VCSELs ($\approx$ 980 nm) the pitch additionally needs to be matched precisely with the wavelength dependent optical characteristics of the DOE (see Fig.\ref{fig1}(a))\,\cite{Brunner.2015}\cite{Maktoobi.2020}. Furthermore, the design of the upper p-contacts and their connections to larger signal pads (for wirebonding to external electrical drivers) facilitates individual addressability of each laser within the array to allow for bias current induced fine-tuning of the emission wavelength (see Fig.~\ref{fig2}(b)). 

The VCSELs are based on a semiconductor heterostructure consisting of a microresonator with 20.5 (37)  Al$_{0.9}$Ga$_{0.1}$As/GaAs  $\lambda/4$ optically thick mirror  pairs with compositionally linearly  graded interfaces in the upper p-doped with C (lower n-doped with Si) distributed Bragg reflectors (DBRs). The central $\lambda/2$ optically thick cavity includes a stack of five rectangular 4\,nm thick GaInAs quantum wells surrounded by ~5\,nm thick GaAsP barrier layers and two 20\,nm thick Al$_{0.98}$Ga$_{0.02}$As layers centered on optical field intensity nodes forming oxide apertures during the VCSEL processing. The fabrication process, shown schematically in Fig.\ref{fig2}(a), starts with the realization of $5\times5$ arrays of ring-shaped p-metal contacts with an outer diameter of 23\,$\mu$m and a width of 6.5\,$\mu$m using UV lithography and electron beam induced metal deposition of 60\,nm Cr, 50\,nm Pt and a capping of 250\,nm Au. Subsequently, the corresponding $5\times5$ laser cavity arrays are patterned using a double-mesa design, which is realized by UV lithography to form SiN hardmasks and inductively coupled plasma reactive ion etching (ICP-RIE) with a slow etching rate, where the latter ensures very smooth and vertical mesa sidewalls. After processing the upper mesas with diameters of 30\,$\mu$m the apertures were oxidized using an oxidation oven with a 420$^{\circ}$C H$_{2}$O+N$_{2}$ atmosphere, a pressure of 50\,mbar and an oxidation time of around 3 hours and 20 minutes. After the selective thermal oxidation and standard wafer cleaning we dry etched the larger lower mesas which have a diameter of 45\,$\mu$m. Next, we deposited and lift-off n-metal ground contacts, which consist of a 40\,nm Ni Layer followed by 100\,nm of the eutectic alloy Au$_{88\%}$Ge$_{12\%}$ and a final capping of a 400\,nm Au layer. To improve the long term stability of the VCSEL arrays, the mesas were then passivated and planarized by a 100\,nm thick SiN layer created by plasma enhanced chemical vapor deposition (PECVD) and spun-on benzo-cyclo-butene (BCB) polymer, respectively. The polymer is reopened selectively using a further UV lithography and a (CF$_{4}$+O$_{2}$)-RIE plasma etching step to again obtain access to the p- and n-contacts. Finally, fabrication is completed by a further UV lithography and metallization process forming larger signal contact areas to connect the individual VCSEL p-contacts with their intended signal pads. These square-shaped signal pads with side lengths between 100 and 200\,$\mu$m are placed on larger areas of the semiconductor material formed in parallel with the mesa etching. These signal pads are electrically isolated to the semiconductor material by an initially 500\,nm thick SiO layer also deposited by PECVD and shaped by wet-chemical etching using buffered oxide etch (BOE). The thickness of the SiO layer is reduced to a remaining thickness of 200 to 300\,nm during the BCB polymer opening plasma RIE etching process. 

An important aspect of our customized VCSEL design is the slightly elliptical cross-section of the VCSELs which ensures tight polarization alignment which is important for the envisioned RC application. The intentional ellipticity stabilizes the laser polarization and aligns it throughout the array to improve the efficiency of optical injection and mutual diffractive coupling of the VCSELs. The elliptical shape was imposed to the upper (smaller) mesa, which also influences the shape of the oxide aperture that in turn has impact on the optical mode confinement. We applied only moderate elliptical distortions with the long mesa axis being about 0.3\,$\mu$m (1\,\%) larger than the short axis. The goal is to control the polarization axis of the lasers' strong mode without degrading the overall emission properties.

\section{Basic emission properties of a $5\times5$ VCSEL array}\label{ioprop}

In this section we present the emission properties of an exemplary $5\times5$ VCSEL array in terms of micro-electroluminescence ($\mu$EL) emission spectra, the input-output characteristics and the polarization control. The optical characterization of the laser array was performed on a two-needle contact stage integrated into a high-resolution $\mu$EL setup. The two freely movable needles allow us to address individual lasers, whose emission was collected by a microscope objective with a numerical aperture (NA) of 0.4 placed in front of the x-y-movable sample stage. The signal was directed to a spectrometer with 0.75 m focal length equipped with an IR-enhanced Si-CCD camera providing a spectral resolution of 30\,$\mu$eV ($\approx$6\,GHz). Motorized $\lambda$/2-plates and linear filters in the detection path enable polarization resolved measurements. The emission power of individual VCSELs was measured with a power meter placed in the detection path behind the microscope objective. 

\begin{figure*}[!t]
	\begin{center}
		\includegraphics[scale=0.2]{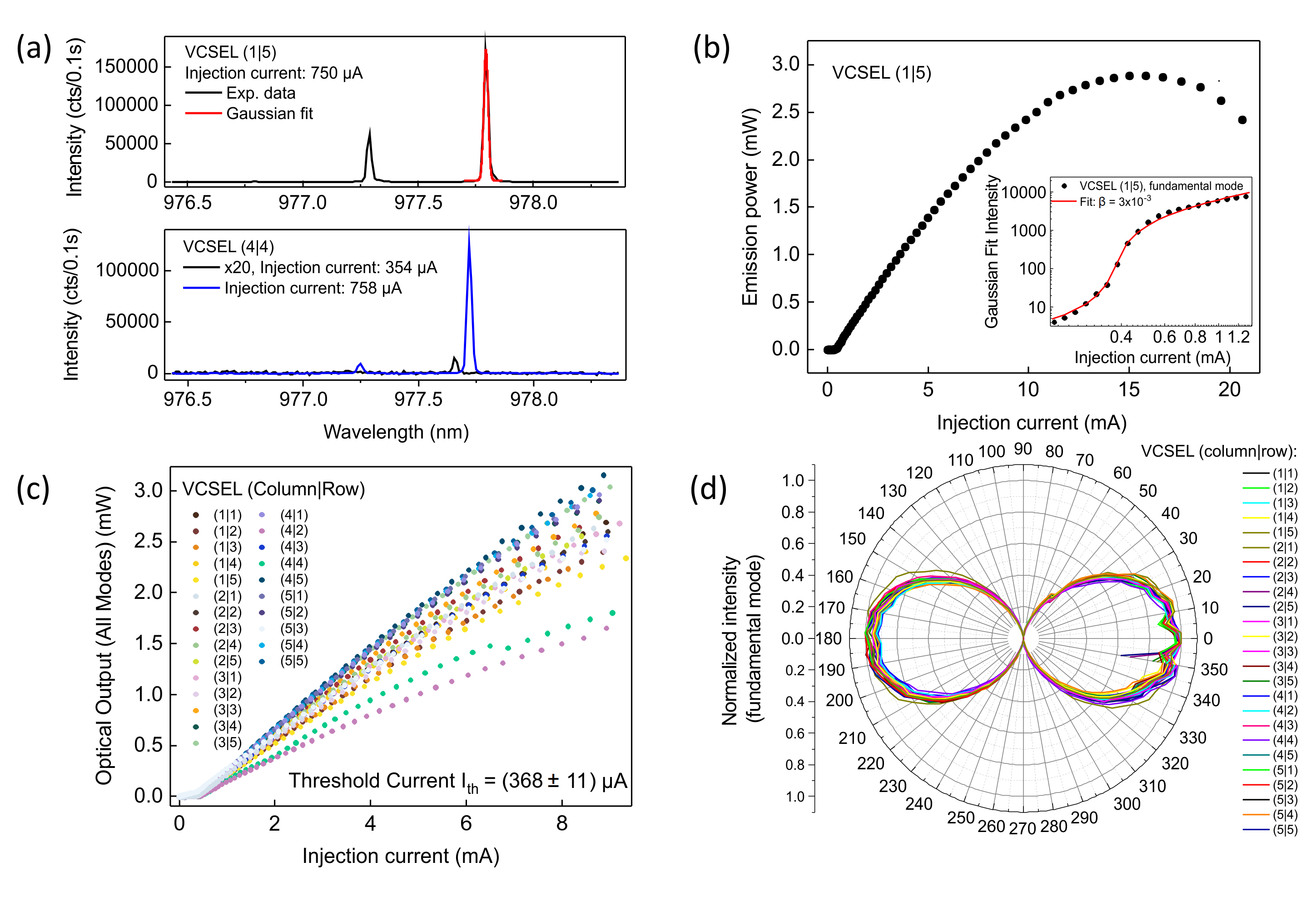}
		\caption{(a) $\mu$EL emission spectra at an injection current of 353, 750 and 758 $\mu$A and (b) emission power (integrated over all cavity modes)
			as function of the current of one laser within the VCSEL array. The inset shows the Gaussian fit intensity of the fundamental mode, fitted with equation from \cite{Reitzenstein.2006}. (c) Emission power of all VCSELs within the $5\times5$ VCSEL array as function of the injection current.  (d) Polarization dependence of the fundamental emission modes of the 25 lasers within the $5 \times 5$ VCSEL array at an injection current of 700$\pm$50\,$\mu$A demonstrating high polarization control due to the slightly elliptical cross-section of the devices.}\label{fig3}
	\end{center}
\end{figure*}

An exemplary $\mu$EL spectrum of VCSEL (column 1| row 5) recorded at an injection current of 750\,$\mu$A (2.2 x $I_{th}$) is presented in Fig.~\ref{fig3}(a). We observe multimode emission with dominating signal from the fundamental transverse cavity mode at 977.8 nm and weaker emission of the next higher lateral mode component at 977.3 nm. This multimode behavior is typical for most of the VCSELs inside this array. Some devices, such as VCSEL (4|4) in Fig.~\ref{fig2}(a), feature single mode emission with side-mode suppression ratios better than 10 dB. Figure \ref{fig3}(b) shows the input-output characteristics of VCSEL (1|5). From the light output curve we infer a threshold current of (352.2$\pm$9.4)\,$\mu$A and in the double-logarithmic presentation of the fundamental mode intensity (inset) we observe a smooth s-shape transition, where we extract a $\beta$-factor of 0.32\,\% by fitting the data with the equation from~\cite{Reitzenstein.2006}. The maximum emission power is observed at about 2.8\,mW (at a pump current of 15 mA), which means that the available optical power range meets the requirements of photonic neural networks \cite{Brunner.2013}\cite{Bueno.2017}.

Moving on to the characteristics of the whole array, we plot the output power as function of the injection current for all 25 devices in linear scale in Fig.~\ref{fig3}(c). All lasers are functional and show very similar laser characteristics with an average threshold current of $(368\pm 11)\,\mathrm{\mu A}$.  Additionally, we observe an average slope efficiency of (0.359$\pm$0.045)\,W/A and maximum output powers of 2 to 3 mW in the chosen bias current range. Only a few lasers feature a lower output power, which correlates with a lower emission wavelength of the fundamental LP01 cavity mode, similar to the shown examples in Fig.~\ref{fig3}(a). The latter is attributed to tighter mode confinement due to a smaller oxide aperture size of the lasers with lower emission powers and indicates a slight spatial inhomogeneity in the oxidation process.

Well-defined polarization properties of the VCSELs are crucial for the target application of the array in photonic RC. We therefore recorded the linear polarization at a bias current of (700$\pm$50)\,$\mu$A and plotted the corresponding angular dependence of the emission in Fig.~\ref{fig3}(d) for all devices within the $5\times5$ array. The almost identical angular dependence of emission of all lasers clearly reflects the polarization alignment achieved by their slight mesa ellipticity of 0.3\,$\mu$m (1\,\%). Indeed, the standard deviation of the polarization orientation has a value of 1.5$^{\circ}$ and the maximum angle difference between two polarizations is only 2.8$^{\circ}$. Noteworthy, in reference measurements we also investigated the polarization dependence of devices with nominally circular mesa shape and smaller ellipticity of 0.15\,$\mu$m. Here, we observed  differences in the polarization orientation of up to 26.9$^{\circ}$ and 15.3$^{\circ}$,  respectively. This comparison highlights the importance of introducing a suitable ellipticity for efficient polarization control of the emission.

Overall, the presented lasing performance of the investigated  $5\times5$ VCSEL array is promising regarding the application in photonic RC. The array features high homogeneity of its emission properties, small threshold currents, suitable output power range and controlled and stable polarization behavior.

\section{Injection locking}

\begin{figure*}[!t]
	\begin{center}
		\includegraphics[scale=0.22]{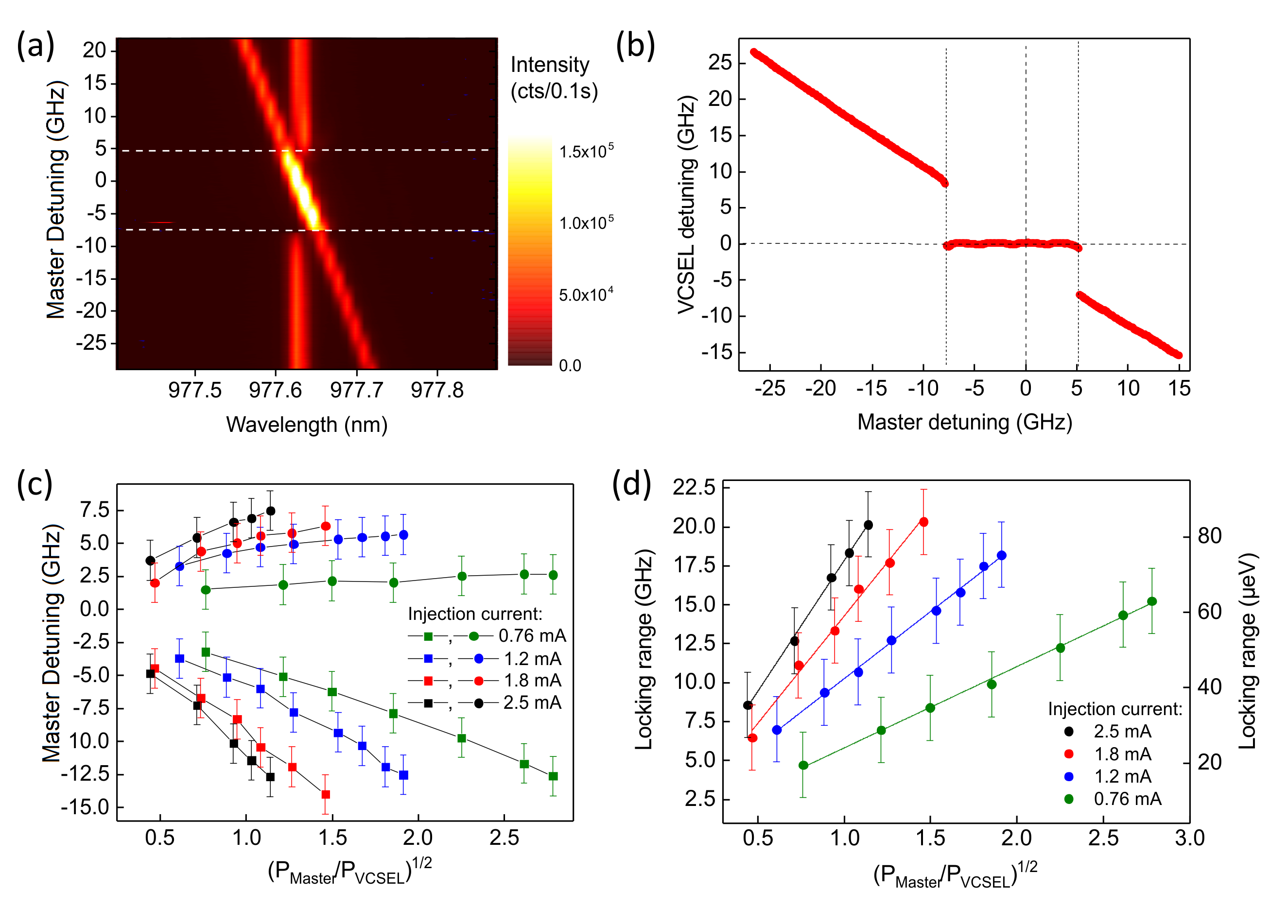}
		\caption{(a) 2D intensity map of VCSEL emission under optical injection by an external tunable laser for a bias current of 1.2\,mA and an injected power of 0.87\,mW. (b) Relative emission frequency of the VCSEL acting as slave laser as a function of the master laser detuning.  (c) Power dependent measurement of the boundaries of the locking area. (d) Injection power dependent locking range for different injection currents.}\label{fig4}
	\end{center}
\end{figure*}

The envisioned implementation of photonic RC is based on the diffractive coupling of the lasers of the VCSEL array and of injection locking to the external information injection laser \cite{Brunner.2013}\cite{Bueno.2017}. In this section we study optical injection locking exemplarily for one laser of the $5\times5$ array. Here, the overarching goal is to determine the power dependent locking range. This is an important figure of merit for the optical coupling and the injection into the semiconductor lasers, because it defines the required spectral homogeneity of the laser array. In fact, efficient diffractive coupling can only be achieved between lasers with a spectral detuning smaller than the mutual injection locking range. To determine the available locking range, optical injection experiments were performed for different powers and power ratios between the external laser acting as master and the VCSELs acting as slave. To determine the effective power ratio at the VCSELs top facet an optical loss of the microscope objective of 30\,\% was taken into account.

In the injection experiments we matched the linear polarization of the master signal to the polarization axis of the strong component of the VCSEL's fundamental mode. Injection locking is explored by spectrally tuning the master laser precisely through resonance with the VCSEL, where we study a tuning range of about 0.2\,nm (60\,GHz). The joint emission of the master laser with an injection power of 870\,$\mu$W and the VCSEL is presented in the 2D intensity plot of Fig.~\ref{fig4}(a) for a bias current of 1.2\,mA (corresponding to a slave power behind the objective of 288\,$\mu$W). We observe pronounced frequency locking to the injection laser, identified by the strongly (by up to a factor of 6 in relation to the master emission) increased joint intensity of the two lasers, in a detuning range of about -8 GHz to about 5 GHz, where the detuning is defined as the difference between the master laser's frequency and the frequency of the considered mode of the VCSEL (emitting at 977.896\,nm). Lineshape fitting of the emission modes allows us to determine the corresponding emission frequencies relative to the solitary slave frequency and to plot them as function of the master-slave detuning. This is depicted in Fig.~\ref{fig4}(b). Using this evaluation we are able to extract the locking range $\Delta$ which is (12.7$\pm$2.1)\,GHz for the data presented in Fig.~\ref{fig4}(b).

The locking range is asymmetric with respect to the slave's solitary frequency, which is a typical phenomenon for semiconductor lasers and can be described by~\cite{Chang.2003}\cite{Schlottmann.2016}:\\
\begin{center}
	\begin{equation}
		- \frac{\nu_{Slave}}{Q}\sqrt{\frac{P_{Master}}{P_{Slave}}}\sqrt{1+\alpha^{2}} < \Delta < \frac{\nu_{Slave}}{Q}\sqrt{\frac{P_{Master}}{P_{Slave}}},  \label{eq1}
	\end{equation}
\end{center}

\noindent \\ where $\nu_{Slave}$ is the resonance frequency and $Q$ is the cavity $Q$-factor of the slave laser. $\alpha$ describes the linewidth enhancement factor and $P_{Master}/P_{Slave}$  is the optical power ratio between master and slave laser. To obtain insight into the underlying physics, we studied the evolution of the locking range's upper and lower boundary as function of $\sqrt{P_{Master}/P_{Slave}}$ for four  different VCSEL bias currents, as shown in Fig.~\ref{fig4}(c). We observe a systematic increase of the locking range and its asymmetry with both, increasing $\sqrt{P_{Master}/P_{Slave}}$ and increasing bias current. Here, the dependence of the locking range on the bias current is attributed to a bias current dependent pre-factor $\nu_{Slave}/Q$ in Eq.\,(\ref{eq1}). While $\nu_{Slave}$ is hardly influenced by the current (apart of a temperature induced red-shift of 142.6 GHz in the relevant current range), a negative correlation between the decreasing effective $Q$-factor and the VCSEL's bias current may be the main reason for the observed current induced increase of the locking range. This current induced decrease of the $Q$-factor can be explained by enhanced optical losses most probably due to heating of the VCSEL's optical cavity.  

We further determined the injection locking range for a wide range of master/slave power ratios from about 0.2 to 8 and plotted the results in Fig.~\ref{fig3}(d), again for four different bias currents. The maximum locking range is about 20 GHz (80 $\mu$eV) for the applied injection powers and bias currents. The experimental values are well approximated by a linear dependence as expected from Eq.(1). Again, the scaling factor, i.e. the slope of the lines, increases with the injection current from 5.2\,GHz to 16.8\,GHz per applied square-root power-ratio. 

The impact of the obtained injection locking results on the application of our VCSEL arrays in photonic RC can be assessed by taking the related laser power budget into account. In the RC concept based on diffractively coupled VCSELs, the signal of the input injection laser with an optical power of typically 100 mW is distributed by a DOE to the individual lasers within the array. Assuming an equal distribution of the available power to all 25 VCSEL we obtain an injection power of 4 mW per laser.  Extrapolating the fitted data in Fig.~\ref{fig4}(d), and under the conservative assumption that only 50\% of this power (i.e. 2 mW) is effectively injected into the laser, we estimate an available locking range of 18.8 - 24.8\,GHz (75 - 102\,$\mu$eV), depending on the bias current. This locking range is an important number for the technological optimization of the VCSEL array's spectral homogeneity which needs to match at least this range to allow for efficient photonic RC~\cite{Bueno.2017}.

\section{Spectral homogeneity of VCSEL array emission}

\begin{figure*}[!t]
	\begin{center}
		\includegraphics[scale=0.7]{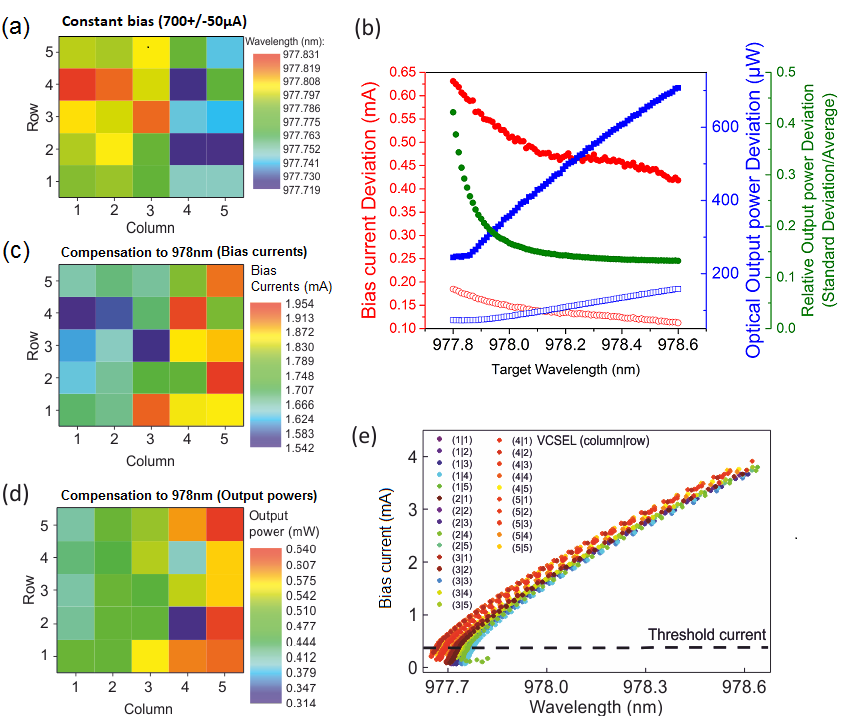}
		\caption{(a) Intrinsic spectral distribution of the VCSEL's fundamental mode wavelength at an constant bias currents of (0.70$\pm$0.05)\,mA. The statistical analysis yields an average wavelength of (977.77 $\pm$ 0.03) nm. (b) Behavior of the deviation in bias currents and optical output dependent on the chosen target tuning wavelength. (c,d) Required bias currents and optical output for each laser for an exemplary wavelength of  978\,nm. The average bias current in (b) is (1.74 $\pm$ 0.12)\,mA and the average power in (c) is (0.50 $\pm$ 0.08)\,mW. (e) Dependence of the bias current on the chosen target tuning wavelength for all VCSELs of the array.}\label{fig5}
	\end{center}
\end{figure*}

After having obtained a detailed characterisation of the input-output and injection locking properties of the fabricated VCSELs, we now turn towards the spectral homogeneity of the $5\times5$ VCSEL array and the spectral tunability of individual VCSELs. As mentioned above, high spectral homogeneity is key to implement high-performance photonic RC based on coupled lasers in an array. Based on the results of the previous section we need to aim at a spectral homogeneity better than $\approx$ 25\,GHz (110 $\mu$eV).

To evaluate the spectral homogeneity of the $5\times5$ VCSEL array, we first consider the case of pumping each laser in the array with the same bias currents of (700$\pm$50)\,$\mu$A. Fig.~\ref{fig5}(a) shows a 2D presentation of the associated, color coded emission wavelengths, with each sub-square being associated with one laser located in the given row and column of the array. We observe a variation of laser wavelengths throughout the array from 977.719\,nm to 977.831\,nm, i.e. with a maximum difference of 0.112\,nm (145.2\,$\mu$eV, 35.12\,GHz), leading to an average wavelength of 977.77\,nm with a standard deviation of only 0.033\,nm (42.4\,$\mu$eV). This high spectral homogeneity is attributed to the high-quality sample processing and to the fact that the processed sample stems from the center of the wafer where the resonance wavelength of the planar microresonator is almost independent of the position. Interestingly, the wavelength variation in the array is low enough that the majority of the lasers in the array are already inside the spectral range needed for efficient injection locking, hence even for uniform bias currents the realized $5\times5$ VCSEL array is already suited for photonic RC. However, per design each laser in the array can be addressed individually so that bias current induced fine-tuning of the emission wavelengths can  further improve the spectral homogeneity.

We demonstrate this fine-tuning by adjusting the wavelength of each laser to the freely chosen target wavelength of 978\,nm. Figure~\ref{fig5}(c) shows the corresponding 2D map in which the bias currents for perfectly uniform emission wavelengths are color coded. Comparing panels (a) and (c) clearly shows that lasers with higher (lower) wavelength need lower (higher) currents to reach the target wavelength. This fine-tuning mode of operation for homogenous emission at 978 nm can be achieved by an average bias current of 1.74\,mA with a standard deviation of 0.122\,mA and a maximum current difference of about 0.412\,mA. However, current induced spectral fine-tuning simultaneously induces an (unwanted) change in the laser output powers. The 2D map of Fig.~\ref{fig5}(d) in which the output power of each VCSEL is color coded, depicts this effect. Here, lasers with higher injection current usually feature higher output power, with some exceptions, like VCSEL (4|4), which has a rather low output power despite a relatively high bias current. We attribute this effect to a lower slope efficiency of this VCSEL because of a slightly smaller oxide aperture. The average output power  of the spectrally tuned array is 0.501\,mW with a standard deviation of 0.082\,mW, which can easily be compensated for by the external optics of the envisaged RC implementation. 

To obtain further insight into the spectral fine-tuning of the individual VCSELs, we recorded the bias current dependent wavelength of each laser's fundamental (LP01) emission mode. The corresponding experimental data for all 25 VCSELs is presented in Fig.~\ref{fig5}(e). The figure essentially shows the bias current needed to achieve a target wavelength in the available range of about 977.8 nm to 978.6 nm for the present array. Here, the wavelength increase above laser threshold with the bias current is due to heating and the corresponding change of the refractive index of the laser devices. Thanks to the high quality of the underlying wafer material and the well-developed device processing both, the input-output characteristics presented in Sec.~\ref{ioprop} as well as the spectral properties are very homogeneous across the $5\times5$ array. To assess and discuss this important point in more detail, we determined the variation of bias currents within the array required to achieve a desired emission wavelength in the possible range of about 977.8 nm to 978.6 nm along with the associated variation of the emission power. Both figures are plotted in Fig.~\ref{fig5} (b) as function of the target wavelength, where open (closed) red circles present the standard deviation (difference between maximum and minimum) of the required bias current range and open (closed) blue squares show the standard deviation (difference between maximum and minimum) of the associated emission power. The standard deviation of bias currents is maximal at small target wavelengths and decreases from about 185 $\mu$A (difference maximum to minimum: 630 $\mu$A) for a target wavelength of 977.8 nm to 110 $\mu$A (420 $\mu$A) for a wavelength of 978.6 nm, while the associated deviation of the emission power increases from 73 $\mu$W (240 $\mu$W) to 160 $\mu$W (700 $\mu$W). These numbers show that each wavelength in the given range can be reached by adjusting the bias current of the individual laser, while keeping the resulting variation of the emission power in an acceptable range well below the  typical injection power. In this regard also the relative deviation of the laser power compared to the average power is of interest and is presented in Fig.~\ref{fig5}(d) as well (green closed circles). In contrast to the absolute values for the associated power, the relative maximum power deviation reaches its maximum of 0.42 at low target wavelengths and is decreasing with longer wavelengths to about 0.13, as the average output power of the array is increasing.

Overall, the measurements of the tuning abilities have shown that the fabricated array allows us to compensate any spectral inhomogenities of the array and align the lasers to achieve a globally injection-locked state. Considering all the discussed data we find that a low target tuning wavelength is the most beneficial regime for operating the VCSEL array as a reservoir system, although the range of choice here is limited by the geometric dimensions of the array pitch and the DOE characteristics \cite{Brunner.2015}\cite{Maktoobi.2020}. On the one hand the main advantage of low target wavelengths is an overall low output power level, which reduces the energy footprint. Considering a bias current of 0.76\,mA the average emission power of the lasers in the array is about (178$\pm$24)\,$\mu$W. Together with the spectral homogeneity in this case of about 42\,$\mu$eV, as estimated above, one can determine with the data from Fig.\,\ref{fig4}(d) a needed injection power ratio of about 3.4 and with that a power consumption for the injection locking for one VCSEL of about 0.609\,mW. Together with typical bias voltages of about 2\,V the total power consumption for one VCSEL would be about 2.1\,mW, and about 45\,mW for the complete array. With this we estimate the energy needed for one induced nonlinear transformation with a value of only 104\,fJ per VCSEL and 2.6\,pJ for the complete array, while considering a modulation bandwidth of the nonlinear transformation of up to 20\,GHz \cite{Fischer.2016}. On the other hand, the major cost for operating at low bias currents and wavelengths is a relatively large distribution of output powers compared to the average value, which leads to a wider distribution of the applied power ratio. While this is indeed a slight complication for achieving the locking state, this disadvantage is partly mitigated by the low increase in the locking range, which makes this regime most suitable for the locking operation.

\section{Summary}

In summary, we presented a photonic hardware platform consisting of a tailored high quality VCSEL laser arrays  developed for the realization of next generation optical neural network concepts, i.e. for reservoir computing (RC). Main challenges for such applications are the optical injection locking operation as well as to achieve the strict requirements for the spectral homogeneity of the laser elements in the array. To meet these goals, we specifically adapted our VCSEL array design. First, we chose an elliptical mesa design which managed to align the polarization of the VCSEL emission to a standard deviation of only 1.52$^{\circ}$. Secondly, a contact design to individually control each laser inside the array was chosen to create a post fabrication wavelength tuning ability via the injection current, where we use the temperature induced resonance shift of the lasers to meet the spectral requirements of optical injection locking and coupling. Overall, the fabricated $5\times5$ VCSEL array show suitable properties for the operation as a reservoir computing system. In addition to the mentioned polarization alignment, the lasers exhibit quite homogeneous lasing transitions around bias currents of (368$\pm$11$)\mu$A. Also the  spectral position of the fundamental LP01 mode emission with an average value of 977.77\,nm with a standard deviation of 0.033\,nm (42.4\,$\mu$eV) give evidence that the array shows excellent intrinsic spectral homogeneity in the case of constant bias currents around (700$\pm$50)\,$\mu$A. Investigations of the injection locking range in dependence of the injected optical power revealed an additional dependence of the locking range on the bias current. Considering realistic injection conditions for RC experiments we found locking ranges between (18.8 - 24.8$\pm$2.1)\,GHz (75-102\,$\mu$eV), depending on the bias current. These locking ranges are suitable to use the array as a reservoir even for constant pumping conditions. Additionally, the individual contact design provides the flexibility to easily spectrally align the lasers inside the array with tuning ranges over a several nm. The consequences of different tuning regimes were discussed and as a result, we suggest a low tuning wavelength, which promises a low consumption per VCSEL of only about 100\,fJ at a bandwidth of 20 GHz. Ongoing work focuses on the application of the fabricated arrays in reservoir computing setups to explore their reservoir computing performance and their potential for further applications in the brain-inspired computing field. Finally, combining with novel integrated and scalable photonic networks based on 3D printed photonic waveguides \cite{Moughames.2019} has high potential to enable fully integrated photonic neural networks based on this technology in the future.

\section{Acknowledgements}

The research leading to these results received funding from the Volkswagen Foundation via the project ''NeuroQNet''. Additionally, the German Research Foundation supported this work via the Collaborative Research Center CRC 787.

%

\section*{References}

\begin{thebibliography}{10}
	\expandafter\ifx\csname url\endcsname\relax
	\def\url#1{{\tt #1}}\fi
	\expandafter\ifx\csname urlprefix\endcsname\relax\def\urlprefix{URL }\fi
	\providecommand{\eprint}[2][]{\url{#2}}
	
	\bibitem{Brunner.2013}
	Brunner D, Soriano M~C, Mirasso C~R and Fischer I 2013 {\em Nature
		Communications\/} {\bf 4} 1364
	
	\bibitem{LeCun.2015}
	LeCun Y, Bengio Y and Hinton G 2015 {\em Nature\/} {\bf 521} 436--444
	
	\bibitem{Silver.2016}
	Silver D, Huang A and Maddison C~e~a 2016 {\em Nature\/} {\bf 529} 484--489
	
	\bibitem{Antonik.2017}
	Antonik P, Haelterman M and Massar S 2017 {\em Cognitive Computation\/} {\bf 9}
	297--306
	
	\bibitem{Wang.2020}
	Wang L, Cho W and Yoon K~J 2020 {\em IEEE Robotics and Automation Letters\/}
	{\bf 5} 1421--1428
	
	\bibitem{Wei.2020}
	Wei J, He J, Zhou Y, Chen K, Tang Z and Xiong Z 2020 {\em IEEE Transactions on
		Intelligent Transportation Systems\/} {\bf 21} 1572--1583
	
	\bibitem{Li.2020}
	Li S, Chen J and Xiang J 2020 {\em Neural Computing and Applications\/} {\bf
		32} 2037--2053
	
	\bibitem{Li.2012}
	Li D, Han M and Wang J 2012 {\em IEEE transactions on neural networks and
		learning systems\/} {\bf 23} 787--799
	
	\bibitem{Yu.2020}
	Yu P and Yan X 2020 {\em Neural Computing and Applications\/} {\bf 32}
	1609--1628 ISSN 0941-0643
	
	\bibitem{Bueno.2017}
	Bueno J, Brunner D, Soriano M~C and Fischer I 2017 {\em Optics Express\/} {\bf
		25} 2401--2412
	
	\bibitem{Hamerly2019}
	Hamerly R, Bernstein L, Sludds A, Solja\ifmmode \check{c}\else
	\v{c}\fi{}i\ifmmode~\acute{c}\else \'{c}\fi{} M and Englund D 2019 {\em
		Physical Review X\/} {\bf 9} 021032
	
	\bibitem{Heuser.2020}
	Heuser T, Große J, Holzinger S, Sommer M~M and Reitzenstein S 2020 {\em IEEE
		Journal of Selected Topics in Quantum Electronics\/} {\bf 26} 1--9
	
	\bibitem{Larger2012}
	Larger L, Soriano M~C, Brunner D, Appeltant L, Gutierrez J~M, Pesquera L,
	Mirasso C~R and Fischer I 2012 {\em Optics Express\/} {\bf 20} 3241--9
	
	\bibitem{Paquot2012}
	Paquot Y, Duport F, Smerieri A, Dambre J, Schrauwen B, Haelterman M and Massar
	S 2012 {\em Scientific reports\/} {\bf 2} 287
	
	\bibitem{Dejonckheere2014}
	Dejonckheere A, Duport F, Smerieri A, Fang L, Oudar J~L, Haelterman M and
	Massar S 2014 {\em Optics Express\/} {\bf 22} 10868
	
	\bibitem{Miscuglio2018}
	Miscuglio M, Mehrabian A, Hu Z, Azzam S~I, George J, Kildishev A~V, Pelton M
	and Sorger V~J 2018 {\em Optical Materials Express\/} {\bf 8} 3851
	
	\bibitem{Fischer.2016}
	{Fischer, I and Bueno, J and Brunner, D and Soriano, M C and Mirasso, C} 2016
	Photonic reservoir computing for ultra-fast information processing using
	semiconductor lasers {\em ECOC 2016\/} vol 42nd European Conference on
	Optical Communication (VDE) pp 1--3
	
	\bibitem{Brunner.2015}
	Brunner D and Fischer I 2015 {\em Optics Letters\/} {\bf 40} 3854--3857
	
	\bibitem{Bueno.2018}
	Bueno J, Maktoobi S, Froehly L, Fischer I, Jacquot M, Larger L and Brunner D
	2018 {\em Optica\/} {\bf 5} 756
	
	\bibitem{Lin2018}
	Lin X, Rivenson Y, Yardimci N~T, Veli M, Jarrahi M and Ozcan A 2018 {\em
		Science\/} {\bf 361} 1004--1008
	
	\bibitem{Shen2016}
	Shen Y, Harris N~C, Skirlo S, Prabhu M, Baehr-Jones T, Hochberg M, Sun X, Zhao
	S, Larochelle H, Englund D and Soljacic M 2017 {\em Nature Photonics\/} {\bf
		11} 441--446
	
	\bibitem{Pai2019}
	Pai S, Bartlett B, Solgaard O and Miller D~A~B 2019 {\em Physical Review
		Applied\/} {\bf 11} 064044
	
	\bibitem{Moughames.2019}
	Moughames J, Porte X, Thiel M, Ulliac G, Jacquot M, Larger L, Kadic M and
	Brunner D Three dimensional waveguide-interconnects for scalable integration
	of photonic neural networks
	
	\bibitem{Khoram2019}
	Khoram E, Chen A, Liu D, Ying L, Wang Q, Yuan M and Yu Z 2019 {\em Photonics
		Research\/} {\bf 7} 823
	
	\bibitem{Jaeger.2004}
	Jaeger H and Haas H 2004 {\em Science (New York, N.Y.)\/} {\bf 304} 78--80
	
	\bibitem{Dion.2018}
	Dion G, Mejaouri S and Sylvestre J 2018 {\em Journal of Applied Physics\/} {\bf
		124} 152132
	
	\bibitem{Yi.2016}
	Yi Y, Liao Y, Wang B, Fu X, Shen F, Hou H and Liu L 2016 {\em Microprocessors
		and Microsystems\/} {\bf 46} 175--183
	
	\bibitem{vanderSande.2017}
	{van der Sande} G, Brunner D and Soriano M~C 2017 {\em Nanophotonics\/} {\bf 6}
	561--576
	
	\bibitem{Maktoobi.2020}
	Maktoobi S, Froehly L, Andreoli L, Porte X, Jacquot M, Larger L and Brunner D
	2020 {\em IEEE Journal of Selected Topics in Quantum Electronics\/} {\bf 26}
	1--8
	
	\bibitem{Andreoli.2020}
	Andreoli L, Porte X, Chr{\'e}tien S, Jacquot M, Larger L and Brunner D Boolean
	learning under noise-perturbations in hardware neural networks
	
	\bibitem{Reitzenstein.2006}
	Reitzenstein S, Bazhenov A, Gorbunov A, Hofmann C, M{\"u}nch S, L{\"o}ffler A,
	Kamp M, Reithmaier J~P, {v d Kulakovskii} and Forchel A 2006 {\em Applied
		Physics Letters\/} {\bf 89} 051107
	
	\bibitem{Chang.2003}
	Chang C~H, Chrostowski L and Chang-Hasnain C~J 2003 {\em IEEE Journal of
		Selected Topics in Quantum Electronics\/} {\bf 9} 1386--1393
	
	\bibitem{Schlottmann.2016}
	Schlottmann E, Holzinger S, Lingnau B, L{\"u}dge K, Schneider C, Kamp M,
	H{\"o}fling S, Wolters J and Reitzenstein S 2016 {\em Physical Review
		Applied\/} {\bf 6}
	
\end{thebibliography}

\providecommand{\newblock}{}

\end{document}